\documentclass[12pt, twocolumn]{scrartcl}

\usepackage{graphicx, nicefrac, siunitx}
\usepackage[british]{babel}
\usepackage{csquotes, xpatch} 
\usepackage[style=nature, sorting=none,maxnames=3, giveninits=true, uniquename=init, sorting=none, url=false,isbn=false,eprint=false,texencoding=utf8,bibencoding=utf8, autocite=superscript, backend=bibtex8]{biblatex}
\usepackage{amsmath, amssymb}
\usepackage{blindtext}
\usepackage{setspace}
\usepackage{authblk}
\usepackage{mathtools}
\usepackage{color}
\usepackage{xcolor}
\usepackage{lmodern}
\usepackage[bookmarks=false]{hyperref}
\bibliography{all_references.bib}

\title{Direct Bound-Electron \textit{g} factor Difference Measurement with Coupled Ions}
\author[1]{Tim Sailer}
\author[1]{Vincent Debierre}
\author[1]{Zolt\'an Harman}
\author[1]{Fabian Heiße}
\author[1]{Charlotte König}
\author[1]{Jonathan Morgner}
\author[1]{Bingsheng Tu}
\author[2,3]{Andrey V. Volotka}
\author[1]{Christoph H. Keitel}
\author[1]{Klaus Blaum}
\author[1]{Sven Sturm}

\affil[1]{Max-Planck-Institut für Kernphysik, Heidelberg, Germany}
\affil[2]{Department of Physics and Engineering, ITMO University, St. Petersburg, Russia}
\affil[3]{Helmholtz–Institut Jena, Jena, Germany}
\date{}

\usepackage[a4paper, left=1.5cm,right=1.5cm,top=1.5cm,bottom=1.5cm,includeheadfoot,includefoot]{geometry} 
\begin{document}
\newcommand\Tstrut{\rule{0pt}{2.6ex}}         
\newcommand\Bstrut{\rule[-0.9ex]{0pt}{0pt}}   
\onecolumn
\normalsize
\maketitle
The quantum electrodynamic (QED) description of light-and-matter interaction is one of the most fundamental theories of physics and has been shown to be in excellent agreement with experimental results \cite{Hanneke2008, Parthey2011, Sturm2011a, Sturm2014, Parker2018, Morel2020}. Specifically, measurements of the electronic magnetic moment (or \textit{g} factor) of highly charged ions (HCI) in Penning traps can provide a stringent probe for QED, testing the Standard model in the strongest electromagnetic fields \cite{Sturm2019}.
When studying the difference of isotopes, even the intricate effects stemming from the nucleus can be resolved and tested as, due to the identical electron configuration, many common QED contributions do not have to be considered.
Experimentally however, this becomes quickly limited, particularly by the precision of the ion masses or the achievable magnetic field stability \cite{Kohler2016}.\\
Here we report on a novel measurement technique that overcomes both of these limitations by co-trapping two HCIs in a Penning trap and measuring the difference of their \textit{g} factors directly. The resulting correlation of magnetic field fluctuations leads to drastically higher precision. We use a dual Ramsey-type measurement scheme with the ions locked on a common magnetron orbit \cite{Rainville2004}, separated by only a few hundred micrometres, to extract the coherent spin precession frequency difference. We have measured the isotopic shift of the bound electron \textit{g} factor of the neon isotopes of $^{20}$Ne$^{9+}$ and $^{22}$Ne$^{9+}$ to 0.56 parts-per-trillion (\num{5.6e-13}) precision relative to their \textit{g} factors, which is an improvement of more than two orders of magnitude compared to state-of-the-art techniques \cite{Kohler2016}. This resolves the QED contribution to the nuclear recoil for the very first time and accurately validates the corresponding theory. Furthermore, the agreement with theory allows setting constraints for a fifth-force, resulting from Higgs-portal-type dark-matter interactions \cite{Debierre2020}.
\newpage
\twocolumn
The theory of QED describes the interaction of charged particles with other fields and the vacuum surrounding them. State-of-the-art calculations of these effects allow for stringent tests of fundamental physics, the search of physics beyond the Standard Model or the determination of fundamental constants \cite{Parthey2011, Sturm2011a, Sturm2014, Parker2018, Morel2020}. One quantity which can be used to perform such tests, is the magnetic moment of an electron bound to a nucleus, expressed by the Landé or \textit{g} factor in terms of the Bohr magneton. It can be accessed experimentally and is also predicted very precisely by theory. Especially hydrogen-like ions, with only a single electron left, provide a simple bound-state system which allows for testing the Standard Model in the extremely strong electric field of the nucleus. In this case, the \textit{g} factor of a free electron is modified by the properties of the nucleus, foremost the additional electric field, but also parameters such as mass, susceptibility and the charge radius have to be considered. 
However, studying these effects explicitly proves to be difficult, since the QED contributions and their uncertainties are significantly larger than many of the nuclear effects, resulting in limited visibility (see Methods ~\ref{tab:calc}).\\
One idea to overcome this limitation is to compare the \textit{g} factors of similar ions, studying the isotopic shift for example. Here, the common identical contributions and their uncertainties do not have to be considered, emphasizing the differences due to the nucleus. In Table \ref{tab:calc_short}, the theoretical contributions and uncertainties to the individual \textit{g} factors of $^{20}$Ne$^{9+}$ and $^{22}$Ne$^{9+}$ and their differences are summarized.

\begin{table}[!tb]
\begin{tabular}{l l}
\textbf{\textit{g} factor (theory)} \Tstrut &\\
\hline
$^{20}$Ne$^{9+}$ & \phantom{$-$}\num{1.998767277114(117)} \Tstrut\Bstrut \\
$^{22}$Ne$^{9+}$ &	\phantom{$-$}\num{1.998767263640(117)} \Tstrut\Bstrut	\\
\textbf{Difference}\Tstrut & \phantom{$-$}($\times$\num{e-9})\\
\hline
\hspace{2mm} FNS 										& \phantom{$-1$}0.166(11) \\
\hspace{2mm} Recoil, non-QED 				& \phantom{$-$}13.283 \\
\hspace{2mm} Recoil, QED 						& \phantom{$-1$}0.043 \\
\hspace{2mm} Recoil, $\alpha(m/M)$ 	& \phantom{1}$-$0.010 \\
\hspace{2mm} Recoil, $(m/M)^2$			& \phantom{1}$-$0.008 \\
\hline \textbf{$\Delta g$ Total theory}\Tstrut\Bstrut	& \phantom{$-$}13.474(11)$_{\text{FNS}}$ \\
\hline \textbf{$\Delta g$ Experiment}\Tstrut\Bstrut	&\phantom{$-$}13.475\,24(53)$_{\text{stat}}$(99)$_{\text{sys}}$\\
\hline
\end{tabular}
\caption{Contributions to the \textit{g} factor difference of $^{20}$Ne$^{9+}$ and $^{22}$Ne$^{9+}$ and the final experimental result. For the individual contributions see Methods \ref{meth_gdiff}.}
\label{tab:calc_short}
\end{table}

For the calculated difference $\Delta g = g(^{20}$Ne$^{9+})-g(^{22}$Ne$^{9+})$, the absolute uncertainty is decreased by two orders of magnitude compared to the absolute values, which allows resolving and testing the QED contribution to the nuclear recoil. This term takes the quantized size of the momentum exchange between electron and nucleus into consideration and requires a fully relativistic evaluation which goes beyond the Furry picture \cite{Furry1951} and the usual external-field approximation \cite{Malyshev2020}. Understanding and confirming this contribution is essential for future \textit{g} factor measurements of heavier ions or when trying to improve upon the precision of the fine-structure constant $\alpha$ \cite{Yerokhin2016}.
Furthermore, a precise measurement of the isotope shift allows searching for physics beyond the Standard Model, by means of looking for a deviation from the calculated effect.  
The Higgs portal scenario in particular postulates a new scalar boson, the relaxion, of unknown mass $m_\Phi$ as a dark matter candidate which would mediate an interaction between nucleons and electrons. The mixing of such a boson with the Higgs boson, with different coupling strengths $y_e$ and $y_n$ for electrons and nucleons respectively, could potentially be directly observed in the isotopic shift due to the different number of neutrons. Specifically, such a measurement would exhibit a strong sensitivity of the \textit{g} factor difference \cite{Debierre2020} for heavy bosons, with a specific energy range of \SI{20}{\MeV} to \SI{1}{\GeV} due to the close proximity of the electron to the nucleus in a HCI (see Methods \ref{sec:newPhysEnergy}). This relaxion, if found, would also provide a solution to the long-standing electroweak hierarchy problem \cite{Graham2015}.
To explicitly study the isotopic shift with formerly unavailable resolution, we report on the application of a newly developed technique to measure the difference of \textit{g} factors directly. This method depends on coupling two ions as a well controlled ion crystal within the magnetic field of a Penning trap. This way, the ions are close enough to be subject to the identical fluctuations of this magnetic field, which otherwise are a strong limiting factor for the achievable precision.
We performed such a measurement in the \textsc{Alphatrap} setup \cite{Sturm2019}. This apparatus consists of a Penning trap \cite{Brown1986} in a superconducting 4-T magnet, where the trap and all detection electronics are cooled by liquid helium to a temperature of about \SI{4.2}{\kelvin}. By combining the magnetic field $B$ and a suitable electrostatic potential, ions can be stored almost indefinitely, limited only by the vacuum quality. A trapped ion's motion can be parametrized by splitting the trajectory into three independent harmonic oscillations that are related to the free cyclotron frequency $\nu_c = \frac{q_\text{ion}}{2\pi m_\text{ion}}B$, with the ion charge and mass $q_\text{ion}$ and $m_\text{ion}$ respectively, via \cite{Brown1986}:
\begin{equation}
\nu_c^2 = \nu_+^2+\nu_z^2+\nu_-^2.
\label{eq:invar}
\end{equation} 
For this measurement on $^{20}$Ne$^{9+}$ and $^{22}$Ne$^{9+}$, the modified cyclotron frequencies $\nu_+$ amount to roughly \SI{27}{\mega\hertz} and \SI{25}{\mega\hertz}, the axial frequencies (parallel to the magnetic field) $\nu_z$ to about \SI{650}{\kilo\hertz} and \SI{620}{\kilo\hertz} and both magnetron frequencies $\nu_-$ to \SI{8}{\kHz}, respectively. These frequencies can be measured non-destructively via the image currents induced by the oscillating charged particle \cite{Cornell1990, Sturm2011}.
Additionally, the presence of the magnetic field results in an energy splitting $\Delta E = h\nu_L$ of the $m_s = \pm\nicefrac{1}{2}$ electronic spin states with the Larmor frequency $\nu_L = \frac{geB}{4\pi m_{e}}$ amounting to about \SI{112}{\GHz}, with the electron mass and charge $m_e$ and $e$, respectively. The orientation $m_s$ of the spin with respect to the magnetic field can be determined by means of the continuous Stern-Gerlach effect \cite{Dehmelt1986a} in the dedicated analysis trap (AT) (see Fig. \ref{fig:setup}). Here, in addition to the homogeneous magnetic field $B_0$, a quadratic magnetic field gradient or \textit{magnetic bottle} $B(z) = B_0+B_1z+B_2z^2$ with $B_2 \approx $ \SI{45}{\kilo\tesla\per\meter\squared} is produced by a ferromagnetic ring electrode. This exerts an additional spin-dependent force on the ion which results in an instantaneous shift of the axial frequency when a millimetre-wave (MW) photon around $\nu_L$ is absorbed. As this magnetic bottle hinders precise frequency measurements, the spectroscopy is performed in the homogeneous magnetic field \cite{Sturm2019} of the precision trap (PT), where also the cyclotron frequency can be measured simultaneously to the MW excitation. The AT is then solely used for the detection of the spin-state and the separation of the ions. The \textit{g} factor can be extracted from the frequencies \cite{Sturm2014, Kohler2016, Arapoglou2019}
\begin{equation}
g = 2\frac{\nu_L}{\nu_c}\frac{m_e}{m_\text{ion}}\frac{q_\text{ion}}{e}.
\label{eq:gee}
\end{equation}
Consequently, the independently measured ion masses, as well as the electron mass, pose direct limits on the achievable precision of absolute \textit{g} factor measurements. Additionally, the inherent magnetic field fluctuations render it impossible to determine the Larmor frequency coherently on the time-scales required to accurately measure the cyclotron frequency. This limits such measurements statistically to low \num{e-11} relative precision even with several months of measurement time and renders an investigation of the small nuclear effects unpractical.

\subsection*{Coupled ions}

\begin{figure*}[!tb]
    \includegraphics[bb= 0 0 590 285,width=\linewidth]{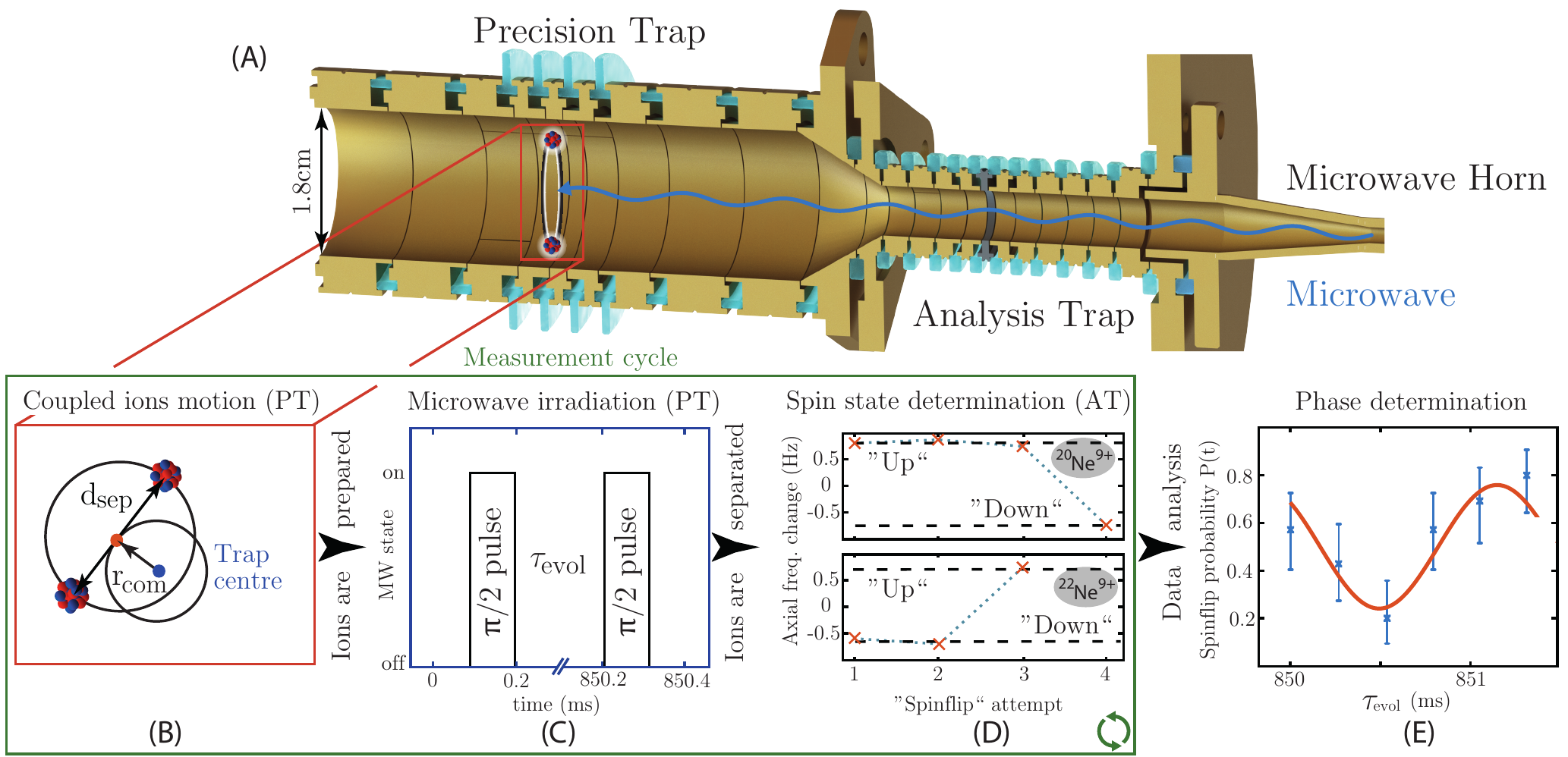}
		\caption{\textbf{(A)} shows the Penning-trap setup, with the coupled ions in the centre of the precision trap. The ions are prepared on a common magnetron orbit, with a separation distance of $d_\text{sep} \approx$ \SI{400}{\micro\meter} and a common mode $r_\text{com} < $ \SI{100}{\micro\meter}. \textbf{(B)} The cyclotron radius of each ion is cooled to about $r_{p} =$ \SI{3}{\um} and the axial amplitude to $r_{z} =$ \SI{18}{\micro\meter} when in thermal equilibrium with the resonator circuit at $T = $ \SI{4.2}{\kelvin}. \textbf{(C)} shows the pulse scheme of MW irradiation. \textbf{(D)} shows the change of axial frequency after each attempt to induce a spin transition. Here, $^{20}$Ne$^{9+}$ was found to be in the ''up'' state, $^{22}$Ne$^{9+}$ in the ''down'' state after the measurement sequence, as can be deduced by the observed change. After several repetitions of such cycles, the coincidental behaviour of the spin-transition rate modulation $P(t)$ is fitted as shown in \textbf{(E)}.}
    \label{fig:setup}
\end{figure*}

To overcome these limitations, we have developed a novel measurement technique, based on the principle of the \textit{Two-Ion-Balance} \cite{Cornell1992, Rainville2004}.
Here, the ions are first prepared separately in the AT to a known electron spin orientation and subsequently merged by placing them in the same potential well of the PT (about \SI{10}{\min}). After cooling the axial motion of the ions individually, they become coupled on a common magnetron orbit due to almost identical frequencies of this mode ($\Delta \nu_- \approx$ \SI{200}{\mHz}), while the axial and modified cyclotron motions remain uncoupled due to their large frequency discrepancy ($\Delta \nu_z \approx$ \SI{20}{\kHz}, $\Delta \nu_+ \approx$ \SI{2.5}{\MHz}). 
The combined motion, as shown in Fig. \ref{fig:setup}, can be parametrized as a superposition of a rotation of both ions with a quasi-static separation distance $d_\text{sep}$ around a common guiding centre on a radius $r_\text{com}$ and a rotation of this guiding centre around the trap centre. The coupling interactions have been mathematically described and used for mass comparison measurements by \cite{Rainville2004}. Now, we determine the initial values of $d_\text{sep}$ and $r_\text{com}$ by measuring the axial frequency shift resulting from the Coulomb interaction of the ions, as well as the individual absolute magnetron radii (merging and determining the initial configuration takes about \SI{10}{\min}).
Subsequently, we are able to convert common mode into separation mode radius, (\cite{Thompson2003} see Methods \ref{sec:comm2sep}) as well as directly cool the separation mode by coupling it to the axial mode. This way, we have full control over all modes as the axial and cyclotron modes of both ions can still be addressed individually.
We apply these tools to prepare the ions with a magnetron separation distance $d_\text{sep} \approx$ \SI{400}{\um} and a comparably small common mode radius $r_\text{com}$ (see Methods \ref{sec:commlimit}, about \SI{20}{\min}).
Now, we perform simultaneous Ramsey-type measurements on the electron spins by irradiating a single millimetre-wave $\pi/2$ - pulse (see Methods \ref{sec:rabimeasurement}) for both ions simultaneously. We then wait for the evolution time $\tau_\text{evol}$, during which both magnetic moments are freely precessing with their individual Larmor frequencies and finally irradiate the second $\pi/2$ - pulse (this takes about \SI{5}{\min}, including a determination of $\nu_c$). Subsequently, the ions are separated again (see Methods \ref{sec:B2sep}, duration \SI{10}{\min}). Finally, the cycle is completed by determining and comparing the spin orientation to the initial state for each ion individually in the AT again. This whole process has been fully automatized, requiring about one hour to complete a cycle. In total, we have performed \num{479} cycles for the main measurement as well as \num{174} for the systematic error analysis.
\begin{figure*}[!tb]
\includegraphics[width=\textwidth]{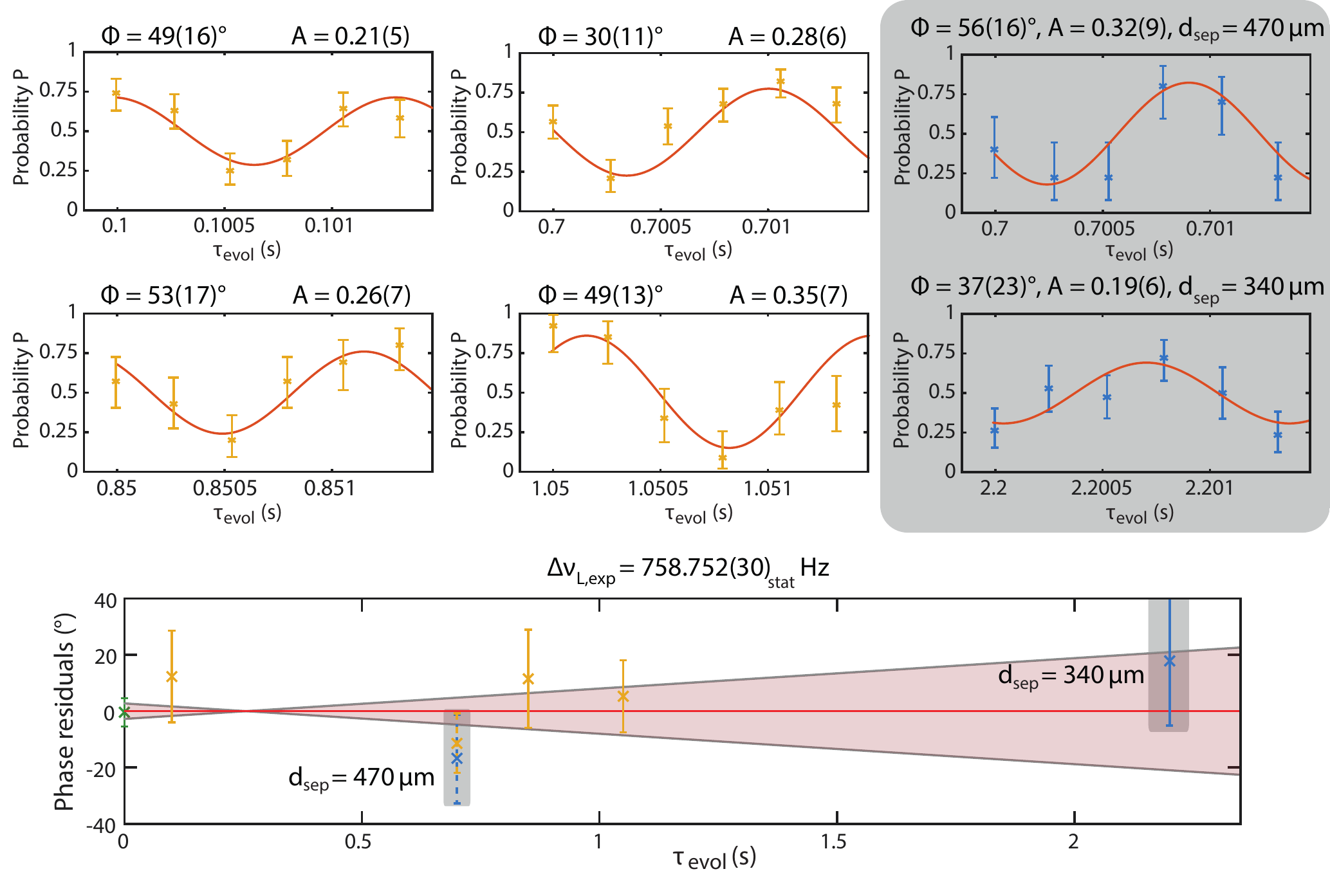}
\caption{The top six figures show the individual measurements. The measurements highlighted in grey do not contribute to the statistical uncertainty of the final result and are only used to confirm and correct for systematic effects. The bottom part shows the residuals with respect to the final frequency, with the $1\sigma$ statistical uncertainty being illustrated in the shaded confidence interval. The initial phase (green) stems from numerical calculation.}
\label{fig:measurement}
\end{figure*}
Due to the fast Larmor precession of \SI{112}{\GHz}, the inherent magnetic field fluctuations lead to decoherence of the applied MW drive frequency with respect to each of the individual spin precessions already after some \SI{10}{\ms} as also observed in Ref. \cite{Britton2016}. However, as the ions are spatially close together, the spins stay coherent with respect to each other as they both experience identical fluctuations.
For each evolution time $\tau_\text{evol}$ of the Ramsey scheme, the individual measurement points are distributed over roughly one period of the difference frequency $\Delta\nu_L \approx $ \SI{758}{\Hz}.
The coherent difference of the precession frequencies can now be extracted by performing a maximum likelihood fit to the correlated spin transition probability $P$. Here, the ions behave identically when their individual spins are in phase, or opposite to each other when the spins are out of phase after the evolution time. We can therefore define
\begin{equation} 
P = p_{1,\text{SF}}\cdot p_{2,\text{SF}}+p_{1,\text{noSF}}\cdot p_{2,\text{noSF}},
\label{eq:prob_def}
\end{equation} 
where $p_{n,\text{SF}}$ and $p_{n,\text{noSF}}$ are the probabilities for ion $n$ to undergo or not undergo a spin-transition, respectively (see Methods \ref{sec:sfprob}). This relation encodes the relative phases of the spins to each other but, due to the loss of coherence with respect to the applied microwave drive, the expected modulation amplitude is only $\pm25\%$. This joint transition probability is therefore directly modulated by the differential phase of the spins and follows the form
\begin{equation}
P(t) = A \cos(2\pi(\nu_{L1} - \nu_{L2})\, t + \phi_{\tau,0}) + \frac{1}{2}, 
\label{eq:fitting}
\end{equation}
encoding the difference of the Larmor frequencies $\Delta \nu_L = \nu_{L1} - \nu_{L2}$.


\section*{Results}

\begin{figure}[!tb]
	\includegraphics[width=\linewidth]{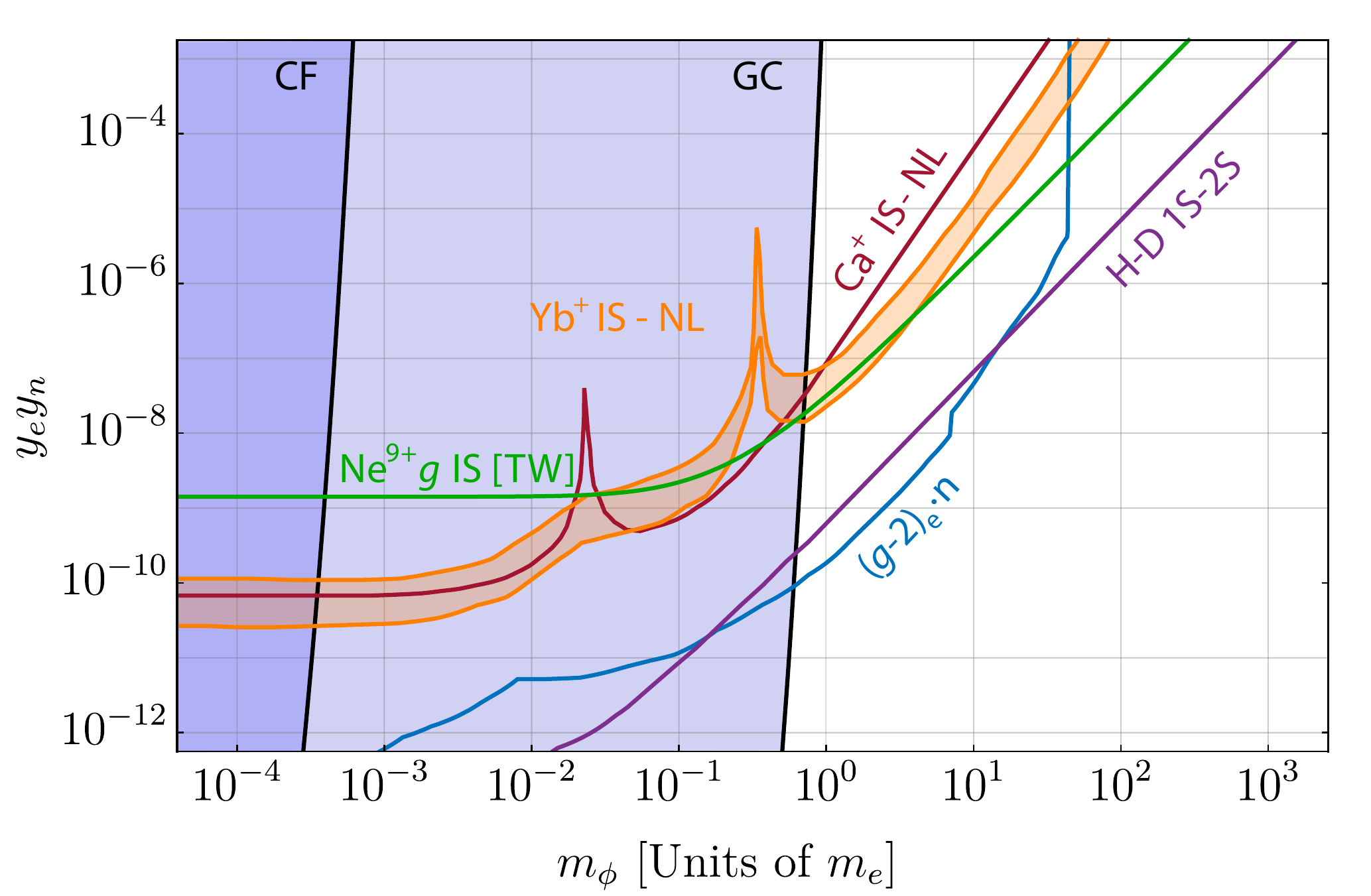}
	\caption{Limits on $y_ey_n$ by different measurements. This measurement [TW] is especially sensitive to comparably large boson masses. For individual references see Methods \ref{sec:newPhysEnergy}.}
	\label{fig:exclusion}
\end{figure}

We have performed measurements for five different sets of evolution times and three different separation distances. Fig. \ref{fig:measurement} shows the modulated probability of a coincidental spin transition occurring for all of these measurements. To extract the Larmor frequency difference, first the total accumulated phase has to be unwrapped. We perform a maximum likelihood fit with a fixed frequency difference, fitting only the phase $\phi_{\tau, 0}$ and amplitude $A$ separately for each evolution time. For all six measurements, the observed amplitude is compatible with a modulation amplitude $A = 25\%$, which confirms the coherent behaviour of the two quantum states for at least up to $\tau_\text{evol} = $ \SI{2.2}{s}, which is more than a factor 20 longer than the coherence time of the individual spins with respect to an external drive. After unwrapping, a linear fit to those phases measured with the separation distance $d_\text{sep} = $ \SI{411(11)}{\um} as well as the calculated initial phase difference (see Methods \ref{sec:init_phase}) is used to determine the frequency difference and the statistical uncertainty. This excludes the two sets with different separation radii, which are used for systematic analysis. Systematic shifts are expected to arise due to the small imbalance of the coupled magnetron motion which is a consequence of the different ion masses. This causes the ions to experience slightly different magnetic fields and alters their individual Larmor frequencies. The two main contributions are firstly this radial imbalance in combination with a residual $B_2$ and secondly a slight shift of the axial equilibrium position caused by a residual deviation from the perfect symmetry of the electrostatic trapping potential, that leads to an unequal change of the Larmor frequencies in the presence of a linear axial $B_1$ field gradient. The combined systematic shift has been evaluated (see Methods \ref{sec:systematic}) to \num{6(5)e-13} relative to the mean Larmor frequency. We specifically stress that our method, while currently experimentally limited by magnetic field inhomogeneities, could be significantly improved by implementing active compensation coils for $B_1$ and $B_2$ \cite{Rau2020}, possibly extending the precision to the \num{e-15} regime.
The bottom plot of Fig. \ref{fig:measurement} shows the residual deviation of each extracted phase with respect to the final frequency difference and uncertainty, corrected for this systematic shift. The grey highlighted data points are for the two measurements performed at a different separation distance, corrected for their expected systematic shift. The agreement of these measurements clearly confirms the systematic correction independently from the calculated correction derived from independent single ion measurements. 
The frequency difference of $\Delta\nu_L = 758.752(30)_{\text{stat}}(56)_{\text{sys}}\hphantom{\cdot}\text{Hz}$, which corresponds to $\Delta g = 13.475\,24(53)_{\text{exp}}(99)_{\text{sys}} \times 10^{-9} $ is in perfect agreement with the theoretical calculation of $\Delta g = $ \num{13.474(11)}$_{\text{FNS}}$, limited in precision solely by the uncertainty of the charge radius difference (finite nuclear size) of the isotopes $\delta\langle r\rangle = $ \SI{0.0530(43)}{\femto\meter} \cite{Angeli2013}. Taking theory as an input instead, our result can thus be applied to improve upon the precision of the charge radius difference by about one order of magnitude $\delta\langle r\rangle = $ \SI{0.0533(4)}{\femto\meter}.  
\\
With the agreement between theory and our result, we are also able set constraints on the scale of the $y_ey_n$ coupling constants which are applied in the new physics (NP) search in the Higgs portal scenario. 
The main difference of our technique compared to the King approach, as for example applied in Ref. \cite{Solaro2020}, can be seen in the range of $50 \hphantom{\cdot} m_e < m_\phi \lesssim 2000 \hphantom{\cdot} m_e $. There, the new physics contribution is cancelled in the large boson mass regime but not in this present approach. Without having to rely upon King plot linearity \cite{Counts2020, Muller2021}, this offers a much more direct approach for the search of new physics. The complete range of applied bounds is shown in Fig. \ref{fig:exclusion}.

\section*{Conclusions \& Outlook}

We have demonstrated and applied our novel method to directly measure a \textit{g} factor difference coherently to unprecedented precision. This is a direct test and validation of the hitherto untested QED contribution to the nuclear recoil and paves the way towards further high-precision measurements on heavier ions where this contribution becomes even larger. 
Alternatively, we are able to improve upon the precision of the charge radius difference by about one order of magnitude using this method, which could be similarly applied to other systems. Additionally, we have applied the result of this single isotopic shift measurement to verify and impose limits on the parameters for the new physics search via the Higgs portal mechanism.

Furthermore, this method provides a crucial step towards accessing the weighted difference of \textit{g} factors \cite{Yerokhin2016, Shabaev2006}, which has the potential to significantly improve upon the precision of the fine-structure constant $\alpha$. Here, the difference between two ions of different nuclear charge $Z$ will have to be measured for both their hydrogen-like $(1s)$ and lithium-like $(2s)$ states using this method. Furthermore, a single absolute \textit{g} factor of low \num{e-11} precision is required when choosing ions of the medium $Z$ range, which has already been shown to be experimentally feasible \cite{Sturm2014}. However, the theoretical calculation of this g-factor has to achieve similar precision which will still require significant work and time.

Finally, the possibility to directly compare matter versus antimatter with highly suppressed systematics should be investigated. This method could possibly be applied to directly compare the anti-proton and $\text{H}^-$  \textit{g} factors. In this case, the Larmor frequency difference would be mostly defined by the electronic shielding of the $\text{H}^-$ ion, which would have to be calculated to similar precision as shown for $^3$He \cite{Rudzinski2009}. Similar to the mass comparison that was already performed \cite{Ulmer2015}, this could enable a direct \textit{g} factor comparison with significantly reduced systematic effects. If a further comparison of proton and positive anti-hydrogen $\overline{\text{H}}^+$ becomes experimentally feasible in the future \cite{Jacob2021}, even the uncertainty of the shielding could be dramatically reduced as well.\\

This work was supported by the Max Planck Society (MPG), the International Max Planck Research School for Quantum Dynamics in Physics, Chemistry and Biology (IMPRS-QD) and the German Research Foundation (DFG) Collaborative Research Centre “SFB 1225 (ISOQUANT)”. This work comprises parts of the Ph.D. thesis work of C.K., T.S. and J.M. to be submitted to Heidelberg University, Germany. This project has received funding from the European Research Council (ERC) under the European Union’s Horizon 2020 research and innovation program under Grant Agreement No. 832848—FunI. Furthermore, we acknowledge funding and support by the Max Planck PTB RIKEN Center for Time, Constants, and Fundamental Symmetries. A.V.V. acknowledges financial support by the Government of the Russian Federation
through the ITMO Fellowship and Professorship Program.

\newpage
\onecolumn
\printbibliography[notkeyword=Methods_only]
\section{Methods}
\subsection{Fitting of Larmor frequency difference}
\label{sec:sfprob}
To derive the fitting function of the correlated spin behaviour of the two ions, we first assume that both ions have been prepared initially in the \textit{spin-down} state, indicated as $\downarrow$. The probability to find each ion individually in the \textit{spin-up} state ($\uparrow$), then follows the probability of a Rabi oscillation with the frequency of the difference between the ion's Larmor frequency $\omega_{L1}$ or $\omega_{L2}$, respectively, and the common microwave drive frequency $\omega_{D}$. 
The probability to find both ions after the measurement sequence in the \textit{spin-up} state follows as

\begin{align}
	P(\uparrow, \uparrow) & = \cos\left(\frac{1}{2}(\omega_{L1}-\omega_D)\tau_\text{evol}\right)^2\cdot \cos\left(\frac{1}{2}(\omega_{L2}-\omega_D)\tau_\text{evol}\right)^2
	\\
	& = \left[\frac{1}{2}\left(\cos\left(\frac{1}{2}(\omega_{L1}-\omega_{L2})\tau_\text{evol}\right)+\cos\left(\frac{1}{2}(\omega_{L1}+\omega_{L2}-2\omega_{D})\tau_\text{evol}\right)\right)\right]^2.
\end{align}

Similarly, the probability to find both ions in the \textit{spin-down} state can be written as

\begin{align}
	P(\downarrow, \downarrow) & = \sin\left(\frac{1}{2}(\omega_{L1}-\omega_D)\tau_\text{evol}\right)^2\cdot \sin\left(\frac{1}{2}(\omega_{L2}-\omega_D)\tau_\text{evol}\right)^2
	\\
	& = \left[\frac{1}{2}\left(\cos\left(\frac{1}{2}(\omega_{L1}-\omega_{L2})\tau_\text{evol}\right)-\cos\left(\frac{1}{2}(\omega_{L1}+\omega_{L2}-2\omega_{D})\tau_\text{evol}\right)\right)\right]^2.
\end{align}

\begin{figure}%
\includegraphics[width=\columnwidth]{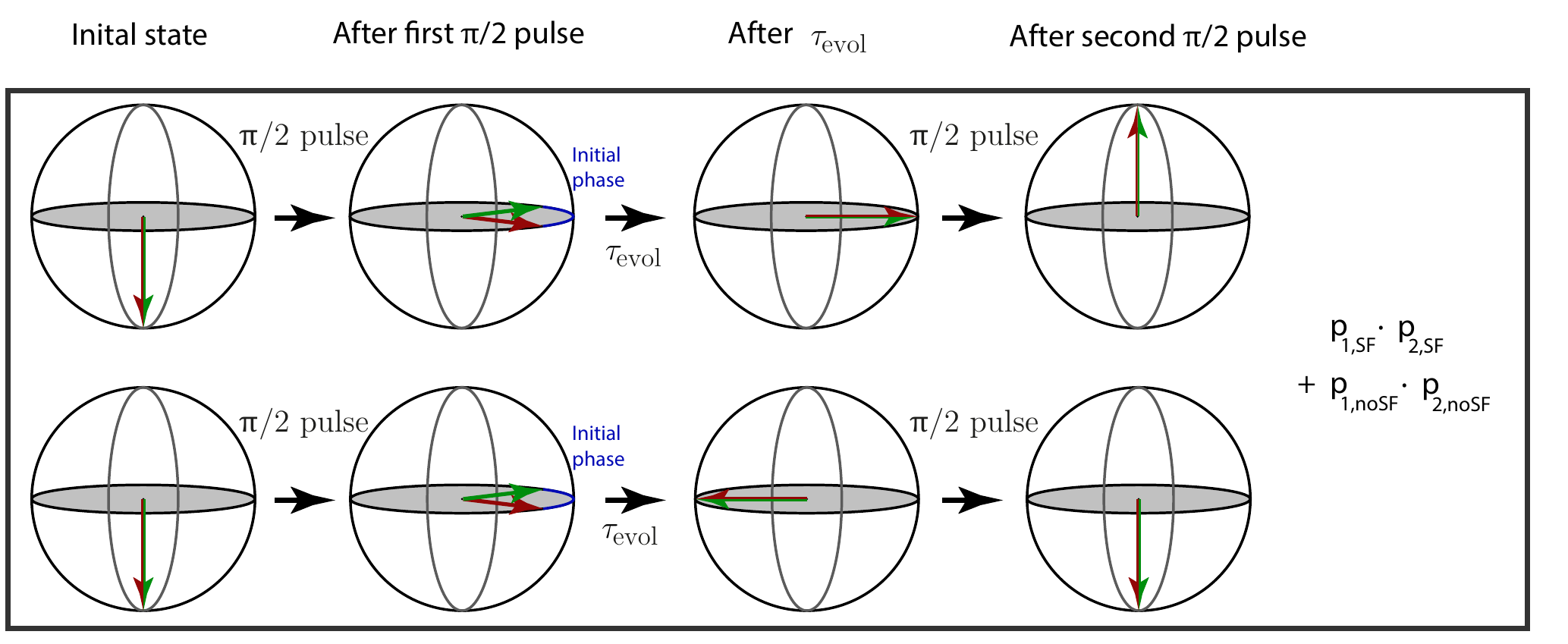}%
\caption{Illustration of two of the possible outcomes of the measurement for an initial configuration with both ions in \textit{spin-down} state. The decoherence of the spins with respect to the microwave drive leads to a reduction in visibility, which is not included in this illustration.}%
\label{fig_SFcases}%
\end{figure}

Both cases have to be considered, as we cannot perform a coherent measurement of the individual Larmor frequencies with respect to the microwave drive, the information about the individual spins is only encoded in the common behaviour.
Therefore, we have to look at the combined probability of both ions either ending up both in the \textit{spin-up} or \textit{spin-down} state (case 1, see Fig. \ref{fig_SFcases}), or the complimentary case, where the two spins behave differently, with one ion in the \textit{spin-up} state, the other ending in the \textit{spin-down} state. The joint probability is given by

\begin{align}
\begin{split}
	P(t) = P(\downarrow, \downarrow) + P(\uparrow, \uparrow) & = \frac{1}{2}\cos\left(\frac{1}{2}(\omega_{L1}-\omega_{L2})\,t\right)^2
	\\
	& \quad + \frac{1}{2}\underbrace{\cos\left(\frac{1}{2}(\omega_{L1}+\omega_{L2}-2\omega_{D})\,t\right)^2}_{\text{1/2}}
	\label{eq:sfprobavg}
	\end{split}
	\\
	& = \frac{1}{4}\cos((\omega_{L1}-\omega_{L2})\,t)+\frac{1}{2},
	\label{eq:sfprob}
\end{align}
where, due to the loss of coherence with respect to the drive frequency, the second term in equation \ref{eq:sfprobavg} averages to \nicefrac{1}{2}. The same formula can be derived for any known initial spin configuration.

\subsection{Calculation of the initial phase}
\label{sec:init_phase}
As our method relies on a single external drive for this specific measurement, used to drive both spins simultaneously, the drive has to be applied at the median Larmor frequency. This results in an additional phase difference that is acquired during the \nicefrac{$\pi$}{2} - pulses. We have determined this phase to be $\Phi_\text{init} = $\SI{35.8(50)}{\degree} using a numerical simulation. Here, we use the knowledge of the Rabi frequency as well as the uncertainty of the magnetic field determination, which leads to an effective jitter of the microwave drive from cycle to cycle. The simulation is performed for different evolution times, extrapolating to the phase that would be measured for zero evolution time. Although the phase that we can extract from the measured data as a cross-check is consistent with this prediction, we still assign an uncertainty of $\pm$ \SI{5}{\degree} to the simulation.

\newpage
\subsection{Combined systematic shifts on $\Delta$\textit{g} of coupled ions with uncertainty of cancellation}
\label{sec:systematic}
Here, we evaluate the total systematic shift and its uncertainty for this method, specifically for the measurement case of $^{20}$Ne$^{9+}$ and $^{22}$Ne$^{9+}$. For this approach, we consider only a separation distance and no common mode. For small common mode radii $r_{\text{com}} \leq$ \SI{100}{\um} which we give as an upper limit, the systematic effects discussed here are actually further reduced \cite{Thompson2003}.
We have to consider multiple individual measurements performed with single ions to characterize these frequency shifts and experimental parameters. More explanation on the methods used can be found in \cite{Sturm2019}, the individual frequency shifts are derived in \cite{Ketter2014}. We define our electric potential, and specifically the coefficients $C_n$ as
\begin{equation}
\Phi(r,\theta) = \frac{V_r}{2}\sum_{n=0}^{\infty}{\frac{C_nr^n}{d_\text{char}^n}P_n(\cos(\theta))}, 
\label{eq:el_pot}
\end{equation} 
with applied ring voltage $V_r$, the characteristic trap size $d_\text{char}$ and the Legendre polynomials $P_n$.
The magnetic field inhomogeneities $B_1$ and $B_2$ are defined as
\begin{align}
\vec{B_1} &= B_1\left(z\vec{e_z}-\frac{r}{2}\vec{e_r}\right)\\
\vec{B_2} &= B_2 \left[\left(z^2-\frac{r^2}{2}\right)\vec{e_z} - zr\vec{e_r}\right].
\label{eq:mag_inhom}
\end{align}

First, we consider the two main axial frequency shifts that depend on the magnetron radius of an ion:

\begin{equation}
\frac{\Delta\nu_z}{\nu_z}\bigg|_{C4} = -\frac{3}{2}\frac{C_4}{C_2d_\text{char}^2}r_-^2
\label{eq:C4mag}
\end{equation}
\begin{equation}
\frac{\Delta\nu_z}{\nu_z}\bigg|_{C3} = \frac{9}{8}\frac{C_3^2}{C_2^2d_\text{char}^2}r_-^2 .
\label{eq:C3mag}
\end{equation}

If the shift of $\nu_z$ is measured to be zero for any radius $r_-$, these two shifts cancel and we can conclude that $C_4 = \frac{3}{4}\frac{C_3^2}{C_2}$.
As it is typically not feasible to tune this for arbitrary radii, especially since higher orders will have to be considered as well for larger radii, we allow a residual $\eta_{el,r_-}$, which includes both, the residual observed shift as well as all neglected smaller contributions. This is a relative uncertainty, scaling with $r^2.$

\begin{equation}
\begin{aligned}
\frac{\Delta\nu_z}{\nu_z}\bigg|_{el} &= \frac{9}{8}\frac{C_3^2}{C_2^2d_\text{char}^2}r_-^2 - \frac{3}{2}\frac{C_4}{C_2d_\text{char}^2}r_-^2
&= \eta_{el,r_-}
\label{eq:C4C3magerr}
\end{aligned}
\end{equation}

Similarly, we consider all frequency shifts that depend on the cyclotron radius $r_+$ of an ion:
\begin{equation}
\frac{\Delta\nu_z}{\nu_z}\bigg|_{C4} = -\frac{3}{2}\frac{C_4}{C_2d_\text{char}^2}r_+^2
\label{eq:C4cyc}
\end{equation}
\begin{equation}
\frac{\Delta\nu_z}{\nu_z}\bigg|_{C3} = \frac{9}{8}\frac{C_3^2}{C_2^2d_\text{char}^2}r_+^2 .
\label{eq:C3cyc}
\end{equation}

The electrostatic contributions are identical to those for the magnetron mode, and per assumption above will also combine to the same $\eta_{el,r_+}$, scaling with the cyclotron radius. However, we have to consider the additional terms that stem from the magnetic field inhomogeneities which are sizeable in this mode due to the significantly higher frequency: 

\begin{equation}
\begin{aligned}
\frac{\Delta\nu_z}{\nu_z}\bigg|_{B_2} &= \frac{B_2}{4B_0}\frac{\nu_++\nu_-}{\nu_+\nu_-}\nu_+r_+^2 \\
&\approx \frac{B_2}{B_0}\frac{\nu_+^2}{2\nu_z^2}r_+^2
\label{eq:B2cyc}
\end{aligned}
\end{equation}

\begin{equation}
\begin{aligned}
\frac{\Delta \nu_z}{\nu_z}\bigg|_{B_1} &= -\frac{3 B_1 C_3 \nu_c\nu_+}{4 B_0 C_2 d_\text{char}\nu_z^2} r_+^2\\
&\approx -\frac{3 B_1 C_3 \nu_+^2}{4 B_0 C_2 d_\text{char}\nu_z^2} r_+^2.\\
\end{aligned}
\label{eq:B1cyc}
\end{equation}

Additionally, for large cyclotron excitations we have to consider the relativistic effect of the mass increase, which also slightly shifts the axial frequency:
\begin{equation}
\frac{\Delta \nu_z}{\nu_z}\bigg|_{rel.} = -\frac{3 B_1 C_3 \nu_c\nu_+}{4 B_0 C_2 d_\text{char}\nu_z^2}r_+^2\\
\label{eq:relcyc}
\end{equation}

The combined shift depending on magnetic inhomogeneities can be expressed as

\begin{equation}
\frac{\Delta \nu_z}{\nu_z}\bigg|_{mag} = \left(\frac{B_2}{B_0}\frac{\nu_+^2}{2\nu_z^2} -\frac{3 B_1 C_3 \nu_+^2}{4 B_0 C_2 d_\text{char}\nu_z^2}\right) r_+^2 = \eta_{mag}.
\label{eq:magshifs}
\end{equation}

While we cannot currently tune these contributions actively (which could be implemented by using active compensation coils \cite{Rau2020}), we can slightly shift the ion from its equilibrium position to a more preferable position along the z-axis to minimize the $B_2$ coefficient. Doing so, we have achieved frequency shifts of $v_z$ close to zero for any cyclotron excitations as well, which means these terms have to cancel as well. We will still allow for another residual error from higher orders, as well as a small residual shift, defined as $\eta_{mag}$. The observed difference of the frequency shift between cyclotron and magnetron excitations $\eta_{mag} + \eta_{el,r_+} +  - \eta_{el,r_-}$ can be used to cancel the identical electric contributions $\eta_{el,r_+}$ and $\eta_{el,r_-}$ when measuring at the same radius. If we solve this combined equation for $C_3$, we are left with only the magnetic field dependent terms $B_1$ and $B_2$, which is what the Larmor frequency difference is sensitive to

\begin{equation}
\begin{aligned}
C_3 &= \frac{2}{3}\frac{B_2C_2d_\text{char}}{B_1}-\underbrace{\frac{4}{3}\frac{B_0C_2d_\text{char}v_z^2}{B_1\nu_+^2r_+^2}\eta_{mag}}_{\xi}\\
&= \frac{2}{3}\frac{B_2C_2d_\text{char}}{B_1}-\xi .
\label{eq:B2dep}
\end{aligned}
\end{equation}

Now, instead of looking at frequency shifts of individual ions, we consider the effects on coupled ions. Due to their mass difference, the coupled state is not perfectly symmetrical but slightly distorted due to the centrifugal force difference. In the case of the neon isotopes, this leads to a deviation of $\delta_{mag} = 0.87\%$, with the definition of $r_1 = d_\text{sep}\frac{(1-\delta_{mag})}{2}$ and $r_2 = d_\text{sep}\frac{(1+\delta_{mag})}{2}$, when choosing ion 1 to be $^{20}$Ne$^{9+}$ and ion 2 as $^{22}$Ne$^{9+}$. Consequently, the frequency difference $\nu_{L_1} - \nu_{L_2}$ will be positive, as the \textit{g} factor (and therefore the Larmor frequency) of $^{20}$Ne$^{9+}$ is larger than for $^{22}$Ne$^{9+}$ We now consider the axial position shift as a function of the slightly different $r_-^2$. This is given by

\begin{equation}
\Delta z = \frac{3}{4}\frac{C_3}{d_\text{char}C_2}r_-^2.
\label{eq:axpos}
\end{equation}

Now we want to express all frequency shifts in terms of $\nu_L$, which is to very good approximation only dependent on the absolute magnetic field, first considering only the effect of $B_1$ and all shifts along the $z$-axis:

\begin{equation}
\frac{\Delta \nu_L}{\nu_L}\bigg|_{B_1}= \Delta z\frac{B_1}{B_0}.
\label{eq:larmorB1}
\end{equation}

The difference of the shift for the individual ions can then be written as
\begin{equation}
\begin{aligned}
\frac{\Delta(\Delta \nu_{L})}{\nu_{L}}\bigg|_{B_1} &= \frac{\Delta\nu_{L_1}-\Delta\nu_{L_2}}{\nu_L} \\
&= \left(\Delta z_1 - \Delta z_2\right)\frac{B_1}{B_0} \\
&= \frac{3}{4}\frac{C_3}{C_2}\frac{B_1}{B_0d_\text{char}}\left(r_1^2-r_2^2\right)\\
&= \left(\frac{1}{2}\frac{B_2}{B_0} - \frac{3}{4}\frac{B_1\xi}{B_0C_2d_\text{char}}\right)(r_1^2-r_2^2)\\
&\eqqcolon \nu_{L,B_1}^{rel}.
\label{eq:larmorB1diff}
\end{aligned}
\end{equation}

We have now the additional uncertainties all summarized in the term scaling with the above-defined factor $\xi$.
The final shift to consider is the same radial difference as mentioned before in the presence of $B_2$. This leads to additional individual shifts in the $\nu_L$ of the ions as

\begin{equation}
\frac{\Delta \nu_L}{\nu_L}\bigg|_{B_2}= \frac{-B_2}{2B_0}r^2.
\label{eq:larmorB2}
\end{equation}

As a relative shift with respect to the measured Larmor frequency difference, this can be written as
\begin{equation}
\begin{aligned}
\frac{\Delta(\nu_{L})}{\Delta\nu_{L}}\bigg|_{B_2} &= \frac{\Delta\nu_{L_1}-\Delta\nu_{L_2}}{\nu_L}\\
&= -\frac{1}{2}\frac{B_2}{B_0}\left(r_1^2 - r_2^2\right)\\
&\eqqcolon \nu_{L,B_2}^{rel}.
\label{eq:larmorB2diff}
\end{aligned}
\end{equation}

Combining these shifts, $\nu_{L,B_2}^{rel}$ and $\nu_{L,B_1}^{rel}$, results in

\begin{equation}
\begin{aligned}
\frac{\Delta(\Delta\nu_{L,tot})}{\nu_{L}} &= \nu_{L,B_1}^{rel} + \nu_{L,B_2}^{rel}\\
&= \left[\frac{1}{2}\frac{B_2}{B_0}-\frac{3}{4}\frac{B_1\xi}{B_0C_2d_\text{char}}-\frac{1}{2}\frac{B_2}{B_0} \right]\left(r_1^2-r_2^2\right)\\
&= -\frac{3}{4}\frac{B_1}{B_0C_2d_\text{char}}\xi\left(r_1^2-r_2^2\right)\\
&= -\frac{3}{4}\frac{B_1}{B_0C_2d_\text{char}}\frac{4}{3}\frac{B_0C_2d_\text{char}v_z^2}{B_1\nu_+^2r_+^2}\eta_{mag}\left(r_1^2-r_2^2\right)\\
&= -\frac{v_z^2}{v_+^2}\frac{\eta_{mag}}{r_+^2}\left(r_1^2-r_2^2\right)\\
&= \num{6e-13} .
\label{eq:finalsys}
\end{aligned}
\end{equation}

We find that, in the ideal case where neither magnetron nor cyclotron excitations produce shifts of the measured axial frequency $v_z$, the final difference of the Larmor frequency is also not shifted at all. Here, we use the worst case, with a measured combined relative shift for $\frac{\eta_{mag}}{r_+^2} \approx \frac{\SI{125}{\mHz}}{\SI{560}{\um\squared}}$. This corresponds to a systematic shift of $\frac{\Delta(\Delta\nu_{L,tot})}{\Delta\nu_{L,tot}} = \num{6e-13}$ which we did correct for in the final result. This been confirmed by performing two measurements on different separation distances, of $d_\text{sep} = $ \SI{340}{\um} and $d_\text{sep} = $ \SI{470}{\um}. Both measurements have been in agreement after correcting for their respectively expected systematic shift.
The uncertainty of this correction of \num{5e-13} has been evaluated numerically by combining the uncertainties of $\eta_{mag}$ and the radii intrinsic to its determination, an uncertainty of $\delta_{\text{mag}}$ and the potential of a systematic suppression of the systematic shift by a residual common mode radius.

\subsection{Different axial amplitudes}
The measurement is performed by first thermalizing the $^{20}$Ne$^{9+}$, then increasing the voltage to bring the $^{22}$Ne$^{9+}$ into resonance with the tank circuit. This will slightly decrease the axial amplitude of the $^{20}$Ne$^{9+}$, which nominally has the larger amplitude when cooled to the identical temperature, compared at the same frequency due to its lower mass. The residual difference in amplitude will lead to a further systematic shift in the presence of a $B_2$, which has been evaluated to about \num{3e-14} and can therefore safely be neglected at the current precision.
 
\subsection{Coupling the ions}

After determining the spin orientations of the individual ions, one ion is excited to a magnetron radius $r_m \approx$ \SI{600}{\um}. The ions are now transported into electrodes next to each other, with only a single electrode in between to keep them separated. Subsequently, this electrode is ramped down as quickly as experimentally possible, limited by DC filters to a time constant of \SI{6.8}{\ms} to keep any voltage change adiabatic compared to the axial frequencies of several \SI{10}{\kHz}. The potentials are also optimized to introduce as little axial energy as possible during this mixing.

\subsection{Preparing the ions for measurement}
\label{sec:comm2sep}
\label{sec:commlimit}
After the ions are mixed, both ions are brought into resonance with the tank circuit one at a time by adjusting the voltage to repeatedly cool their axial modes. Once thermalized, the axial frequency is automatically measured and adjusted to the resonance frequency. From the observed shift in axial frequency compared to a single cold ion, the separation distance $d_\text{sep}$ of the ions can already be inferred, without further information about the common mode however. At this point, both ions are cooled in their respective cyclotron motions via sideband coupling \cite{Cornell1990}. The common mode radius $r_\text{com}$ of the coupled ions can be measured by applying a $C_4$ field contribution, causing the axial frequency to become dependent on the magnetron radius. With the amplitudes of the axial and reduced cyclotron motion being small, this frequency shift allows for determining the RMS magnetron radius of each ion. If the common mode is large, the modulation of the magnetron radius, due to slightly different frequencies of separation and common mode, will lead to visible sidebands due to the axial frequency modulation.\\
For small common mode radii, we will simply measure half the separation radius for each ion. In combination with the known separation distance, the common mode radius can now be determined, however due to limited resolution of the axial frequency shift and the quadratic dependency, $r_\text{com} \approx \sqrt{r_{rms}^2-(\frac{1}{2}d_\text{sep})^2}$, a conservative uncertainty after the ion preparation of $r_\text{com} = $ \SI{0(100)}{\um} is assumed. For consistency, we have prepared the ions in the final state and again excited the common mode to a known radius which could be confirmed using this method.
In case of a large initial common mode, we first have to cool it. Unfortunately, addressing it directly is complicated, as the separation mode will always be cooled as well. However, using the method described in \cite{Thompson2003}, we are able to transfer the common mode radius to the separation mode. This requires a non-harmonic trapping field with a sizeable $C_4$, combined with an axial drive during this process. The axial frequency will now be modulated due to the detuning with $C_4$ in combination with the modulated radius due to the common mode. As the ion will only be excited when being close to the drive, we gain access to a radius dependant modulation force, that finally allows the coupling of common and separation mode.\\
Finally, with the common mode thus sufficiently cooled, we directly address the separation mode, cooling it to the desired value. Due to the strong axial frequency change during cooling, scaling with $d_\text{sep}^3$ and typically being in the range of $\Delta\nu_z \approx$ \SI{150}{\Hz}, the final radius cannot be exactly chosen but rather has a distribution that scales with the power of the cooling drive used. Therefore, one can chose to achieve more stable radii at the cost of having to perform more cooling cycles, ultimately increasing the measurement time. We choose a separation distance $d_\text{sep} = $ \SI{411(11)}{\um}, with the uncertainty being the standard deviation of all measurements as an acceptable trade-off between measurement time and final separation distance distribution. Furthermore, while a smaller separation distance directly corresponds to a decreased systematic uncertainty (see Methods \ref{sec:systematic}), the increased axial frequency shift as well as a deteriorating signal quality of the coupled ions result in a practical limit around $d_\text{sep} = $ \SI{300}{\um}.

\subsection{Measurement}

Before irradiating the microwave pulses, the cyclotron frequency is measured via the double dip technique using $^{22}$Ne$^{9+}$. This measurement is required to be accurate to only about \SI{100}{\mHz}, which corresponds to a microwave frequency uncertainty of about \SI{400}{\Hz}, which is neglectable considering a Rabi frequency of over \SI{2}{\kHz} for a spin transition. The microwave pulse is applied at the median of the Larmor frequencies of $^{22}$Ne$^{9+}$ and $^{20}$Ne$^{9+}$ and therefore detuned from each Larmor frequency by about \SI{380}{\Hz}. This detuning is taken into account when calculating the required time for a $\pi/2$ - pulse. 

\subsection*{Separation of ions}
\label{sec:B2sep}
The strong magnetic bottle, or $B_2$ coefficient that is present in the AT gives rise to a force dependent on the magnetic moment of the ion. The main purpose is to allow for spin-flip detection via the continuous Stern-Gerlach effect. Additionally, this $B_2$ can be utilized to create different effective potentials for the ions depending on their individual cyclotron radii $r_+$. These give rise to the magnetic moment $\mu_{cyc}=\pi\nu_+ q_\text{ion} r_{p}^2$, which then results in an additional axial force in the presence of a $B_2$.
To use this effect to separate the coupled ions, one of them is pulsed to $r_{p} \approx $ \SI{800}{\um} at the end of the measurement in the PT. Subsequently, both ions are cooled in their magnetron modes, resulting in a state where one ion is in the centre of the trap at thermal radii for all modes while the other is on the large excited cyclotron radius. We verify this state by measuring the radii of both ions to confirm the successful cooling and excitation. Now, we use a modified ion transport procedure, with the electrode voltages scaled such that the ion with $r_{p} >$ \SI{700}{\um} cannot be transported into the AT but rather is \textit{reflected} by the $B_2$ gradient, whereas the cold ion follows the electrostatic potential of the electrodes. The hot ion is transported back into the PT and can be cooled there, leaving both ions ready to determine their electron spin orientation again, completing a measurement cycle. This separation method has worked flawlessly for over \num{700} attempts.

\subsection*{Rabi measurement}
\label{sec:rabimeasurement}
To determine the required $\pi/2$ - pulse duration, a single ion, in this case $^{22}$Ne$^{9+}$ is used. We determine the spin orientation in the AT, transport to the PT, irradiate a single microwave pulse and check the spin orientation again in the AT. Depending on the pulse duration, the probability of achieving a change of spin orientation follows a Rabi cycle as 
\begin{equation}
\begin{aligned}
P(SF) &= \frac{\Omega_R^2}{\Omega_R^{'2}} \sin^2(\Omega_R^{'}\pi t))\\
\Omega_R^{'} &= \sqrt{(\Omega_R^2+\Delta\Omega_L^2)}.
\label{eq:Rabi}
\end{aligned}
\end{equation} 
 
Here, $\Omega_R$ is the Rabi frequency and $\Delta\Omega_L$ the detuning of microwave drive with respect to the Larmor frequency. With a measured Rabi frequency of $\Omega_R =$ \SI{2465(16)}{\Hz}, we can irradiate the mean Larmor frequency of the two ions, with the difference being about \SI{758}{\Hz}. Thereby, we are able to use a single pulse simultaneously for both ions while accounting for the detuning to achieve a $\pi/2$ - pulse of \SI{101.1}{\us} for both ions simultaneously.

\begin{figure}[!tb]%
\begin{center}
\includegraphics[width=0.5\textwidth]{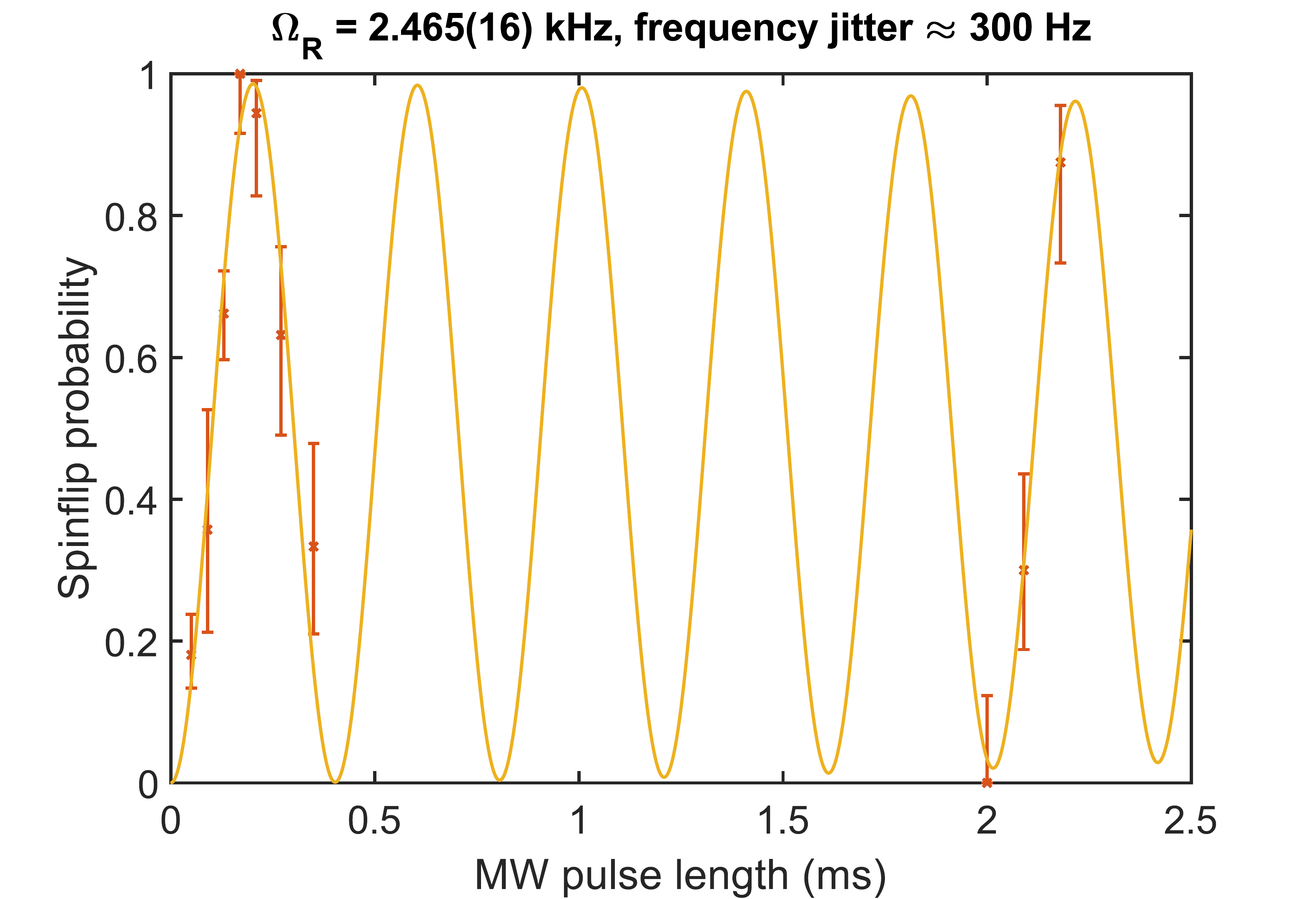}%
\caption{The measured Rabi frequency $\Omega_R$ on a single ion. The probability of inducing a change of spin orientation is modulated by the pulse length of the microwave pulse.}%
\label{fig:Ramsey}%
\end{center}
\end{figure}

\subsection*{Measurement of charge radii differences}
\label{sec:Methods_radii}
We would in principle be able to improve upon any charge radii differences, where this is the limiting factor for the theoretical calculation of \textit{g}. This holds true for most differences between nuclear spin free isotopes, as well as differences between different atoms, provided they are either light enough for theory to be sufficiently precise or close enough in nuclear charge $Z$ such that the corresponding uncertainties are still strongly suppressed.

\subsection{\textit{g} factor calculation}

\label{meth_gdiff}
In Table \ref{tab:calc} the individual contributions to the \textit{g} factors of both ions are shown. The main uncertainty, the higher order two loop QED contribution, is identical for both ions and does cancel in their difference and can therefore be neglected for the uncertainty of $\Delta g$.
\begin{table*}[h]
\begin{tabular*}{\textwidth}{l l l}
& \multicolumn{1}{c}{$^{20}$Ne$^{9+}$} \Tstrut\Bstrut & \multicolumn{1}{c}{$^{22}$Ne$^{9+}$} \Tstrut\Bstrut \\
\hline
Dirac value (point nucleus) 			& \phantom{$-$}\num{1.996445170898(2)}  	& \phantom{$-$}\num{1.996445170898(2)} \\  
Finite nuclear size, FNS    			& \phantom{$-$}\num{0.000 000 004 762(7)} & \phantom{$-$}\num{0.000 000 004 596(12)} \\
QED, one loop $(\alpha)$ 									& \phantom{$-$}\num{0.002325473294(1)}  	& \phantom{$-$}\num{0.002325473294(1)}\\
QED, two loop $(\alpha)^2$ 								& $-$\num{0.000003547780(117)} 						& $-$\num{-0.000003547780(117)} \\
QED, $\geq$three loop $(\alpha)^{3+}$								& \phantom{$-$}\num{0.000000029524} 			& \phantom{$-$}\num{0.000000029524}\\
Nuclear recoil & \\
\hspace{2mm} Non-QED 							& \phantom{$-$}\num{0.000 000 146 093 420}  	& \phantom{$-$}\num{0.000 000 132 810 693} \\   
\hspace{2mm} QED	 								& \phantom{$-$}\num{0.000 000 000 478 954}  	& \phantom{$-$}\num{0.000 000 000 434 499} \\
\hspace{2mm} $(\alpha/\pi)(m/M)$ 	& $-$\num{0.000 000 000 113 2(6)} 							& $-$\num{0.000 000 000 102 9(5)} \\
\hspace{2mm} $(m/M)^2$      			& $-$\num{0.000 000 000 044 1(2)} 							& $-$\num{0.000 000 000 036 5(2)}\\
Hadronic effects									& \phantom{$-$}\num{0.000 000 000 003} 		& \phantom{$-$}\num{0.000 000 000 003} \\
\hline
\textbf{\textit{g} factor Total theory} 										& \phantom{$-$}\num{1.998767277114(117)} \Tstrut\Bstrut  &	\phantom{$-$}\num{1.998767263640(117)}	\Tstrut\Bstrut	\\
\hline
Difference (in \num{e-9}) \Tstrut & &\\
\hspace{2mm} FNS 										& \multicolumn{2}{c}{\phantom{$-$}0.166(11)}\\
\hspace{2mm} Recoil, non-QED 				& \multicolumn{2}{c}{\phantom{$-$}13.283}\\
\hspace{2mm} Recoil, QED 						& \multicolumn{2}{c}{\phantom{$-$}0.043}\\
\hspace{2mm} Recoil, $\alpha(m/M)$ 	& \multicolumn{2}{c}{$-$0.010}\\
\hspace{2mm} Recoil, $(m/M)^2$			& \multicolumn{2}{c}{$-$0.0076}\\
\hspace{2mm} Deformation					 	& \multicolumn{2}{c}{$<0.0001$}\\
\hspace{2mm} Polarisation					 	& \multicolumn{2}{c}{$<0.002$}\\
\hline \textbf{$\Delta g$ Total theory}\Tstrut\Bstrut	& \multicolumn{2}{c}{\phantom{$-$}13.474(11)$_{\text{FNS}}$} \\
\hline \textbf{$\Delta g$ Experiment}\Tstrut\Bstrut	& \multicolumn{2}{c}{\phantom{$-$}13.47524(53)$_{\text{stat}}$(99)$_{\text{sys}}$} \\
\hline
\hline

\end{tabular*}
\caption{Contributions to the calculation of the \textit{g}-factors of $^{20}$Ne$^{9+}$ and $^{22}$Ne$^{9+}$ and their difference and the final experimental result. TW = this work.}
\label{tab:calc}
\end{table*}

\subsection{Setting Limits on new physics}
\label{sec:newPhysEnergy}
Measuring the $g$ factor allows for high-precision access to the properties of very tightly bound electrons, and hence to short-range physics, including potential new physics (NP). Bounds on NP can be set with isotope shift data on the $g$ factor of H-like Ne. The Higgs portal mechanism, in particular, involves the mixing of a new (massive) scalar boson, the relaxion, with the Higgs boson. 
It has been proposed as a solution to the long-standing electroweak hierarchy problem~\cite{Graham2015} with the relaxion as a dark matter candidate~\cite{Banerjee2019}. Constraints on this proposed extension of the Standard Model (SM) can be set with cosmological data, as well as with particle colliders, beam dumps, and also with smaller, high-precision experiments (see, e.g., Ref.~\cite{Frugiuele2017} and references therein).\\
The most common approach in atomic physics is to search for deviations from linearity on experimental isotope shift data in a so-called King plot analysis~\cite{Frugiuele2017,Delaunay2017a,Mikami2017,Berengut2018,Flambaum2018,Solaro2020}, which can be a sign of NP, although nonlinearities can also happen within the SM~\cite{Debierre2020,Flambaum2018,Yerokhin2020,Counts2020}, which limits the bounds which can be set on NP parameters. Here, we present constraints on NP from data on a single isotope pair.

The influence of Higgs Portal relaxions (scalar bosons) on atoms can be expressed~\cite{Frugiuele2017,Mikami2017,Berengut2018,Flambaum2018} by a Yukawa-type potential (often called `fifth force') exerted by the nucleus on the atomic electrons:
\begin{equation} \label{eq:HPPotential}
  V_{\mathrm{HP}}\left(\mathbf{r}\right)=-\hbar c\,\alpha_{\mathrm{HP}}\,A\,\frac{\mathrm{e}^{-\frac{m_\phi c}{\hbar}\left|\mathbf{r}\right|}}{\left|\mathbf{r}\right|},
\end{equation}
where $m_\phi$ is the mass of the scalar boson, $\alpha_{\mathrm{HP}}=y_ey_n/4\pi$ is the Higgs portal coupling constant, with $y_e$ and $y_n$ the coupling of the boson to the electrons and the nucleons, respectively, and $A$ is the nuclear mass number. Yukawa potentials naturally arise when considering hypothetical new forces mediated by massive particles. The corresponding correction to the H-like $g$ factor is given by~\cite{Debierre2020}
\begin{equation} \label{eq:GNP}
  g_{\mathrm{HP}}=-\frac{4}{3}\alpha_{\mathrm{HP}}\,A\,\frac{\left(Z\alpha\right)}{\gamma}\,\left(1+\frac{m_\phi}{2Z\alpha m_e}\right)^{-2\gamma} \times\left[1+2\gamma-\frac{2\gamma}{1+\frac{m_\phi}{2Z\alpha m_e}}\right],
\end{equation}
where $\gamma=\sqrt{1-\left(Z\alpha\right)^2}$. The mass scale of the hypothetical new boson is not known~\cite{Frugiuele2017}, apart from the upper bound $m_\phi<60~\mathrm{GeV}$. In the small boson mass regime $m_\phi\ll Z\alpha m_e$, the contribution to the $g$ factor simplifies to
\begin{equation} \label{eq:GNPLight}
  g_{\mathrm{HP}}=-\frac{4}{3}\alpha_{\mathrm{HP}}\,A\,\frac{\left(Z\alpha\right)}{\gamma}, \quad \text{for}\ m_\phi\ll Z\alpha m_e.
\end{equation}
In the large boson mass regime $m_\phi\gg Z\alpha m_e$, on the other hand, we obtain
\begin{equation} \label{eq:GNPHeavy}
  g_{\mathrm{HP}}=-\frac{4}{3}\alpha_{\mathrm{HP}}\,A\,\frac{\left(Z\alpha\right)\left(1+2\gamma\right)}{\gamma}\left(\frac{m_\phi}{2Z\alpha m_e}\right)^{-2\gamma},\quad \text{for}\ m_\phi\gg Z\alpha m_e.
\end{equation}

We can set bounds on the NP coupling constant by comparing the measured and calculated values of the $g$-factor isotope shift (see Ref.~\cite{Debierre2020}, and also Ref.~\cite{Jaeckel2010} for an implementation of the same idea with transition frequencies in atomic systems). Uncertainties from theory are a source of limitation in this approach. The SM contributions to the isotope shift of the $g$ factor of H-like Ne are given in Table~\ref{tab:calc}, as calculated in this work based on the approaches developed in the indicated references. As can be seen, the largest theoretical uncertainty comes from the leading finite nuclear size correction, and is due to the limited knowledge of nuclear radii (the uncertainty on the finite nuclear size correction due to the choice of the nuclear model is negligible at this level of precision). We note that the standard source for these nuclear radii is data on X-ray transitions in muonic atoms~\cite{Angeli2013}.\\ 
In the NP relaxion scenario, the energy levels of these muonic atoms is also corrected by the relaxion exchange. Another source of root-mean-square charge radii and their differences is optical spectroscopy. The electronic transitions involved are far less sensitive to hypothetical NP than muonic X-ray transitions. The radius difference between \textsuperscript{20}Ne and \textsuperscript{22}Ne extracted from optical spectroscopy~\cite{Ohayon2019} agrees with the one determined from muonic atom data within the respective uncertainties, which shows that NP need not be taken into account to extract nuclear radii from these experiments at their level of precision. To conclude, for our purposes, hypothetical contributions from NP do not interfere with the interpretation of muonic atom data for the extraction of nuclear radii.\\
Taking $\Delta g_{\mathrm{theo}}^{AA'}=1.1\times10^{-11}$ as the theoretical error on the isotope shift, it can be seen from Eq.~(\ref{eq:GNPLight}) that this corresponds to an uncertainty of $\Delta y_ey_n\sim7.1\times10^{-10}$ (and a 95\% bound on $y_ey_n$ twice as large as this) in the small boson mass regime $m_\phi\ll Z\alpha m_e$, which is weaker than the currently most stringent bounds coming from atomic physics (H-D 1S-2S, ~\cite{Delaunay2017}). In the large boson mass regime $m_\phi\gg Z\alpha m_e$, our bound remains weaker, but becomes more competitive, and is more stringent than those of Ref.~\cite{Solaro2020}, thanks to two favourable factors. First, the nuclear charge $Z$ in Eq.~(\ref{eq:GNPHeavy}) is somewhat larger than the screened effective charge perceived by the Ca\textsuperscript{+} valence electron, and larger than the charge of the H nuclei, which also enter the scaling of the bound obtained with these respective ions~\cite{Berengut2018}. Second, when carrying out a King analysis as done in Ref.~\cite{Solaro2020}, one works with two different transition frequencies, and the leading term in the hypothetical NP contribution in the large boson mass regime, which is the equivalent of the r.h.s. of Eq.~(\ref{eq:GNPHeavy}), is cancelled out in the nonlinearity search, due to its proportionality to the leading finite nuclear size correction~\cite{Berengut2018}, leaving the next term, which scales as $\left(m_\phi/\left(2Z\alpha m_e\right)\right)^{-1-2\gamma}$, as the first nonvanishing contribution.\\
In the present case, the $g$ factor of a single electronic state is considered (for a single isotope pair), and this cancellation does not occur. This leads to competitive bounds in the large boson mass regime with the simple $g$ factor isotope shift of H-like ions, as shown in Fig.~\ref{fig:exclusion} (where we used the exact result, Eq.~(\ref{eq:GNP})). We compare our bounds on the coupling constant $y_ey_n=4\pi\alpha_{\mathrm{HP}}$, to the bounds obtained in Refs.~\cite{Solaro2020,Delaunay2017}, through isotope shift measurements in Ca\textsuperscript{+} (Ca$^+$ IS-NL) and H (with nuclear radii extracted from muonic atom spectroscopy), as well as to the bounds obtained through Casimir force (CF) measurements~\cite{Bordag2009}, globular cluster (GC) data~\cite{Redondo2013}, and a combination of neutron scattering and free-electron $g$ factor data ($(g-2)_e\cdot n$ ~\cite{Solaro2020}).\\
We also reproduce the preferred range for the coupling constant obtained in Ref.~\cite{Counts2020}, through isotope shift measurements in Yb\textsuperscript{+}(Yb$^+$ IS-NL). This range was obtained by assuming that the observed King nonlinearity in the experimental isotope shift data is caused by NP. By contrast, all nuclear corrections to the $g$ factor which are relevant at the achieved experimental precision were taken into account in our approach, allowing for an unambiguous interpretation of the experimental data.
\newpage
\printbibliography[keyword=Methods_only]
\end{document}